\documentclass[11pt]{article}
\usepackage{moriond,epsfig}
\usepackage{bm}

\def\be{\begin{equation}}
\def\ee{\end{equation}}
\def\bea{\begin{eqnarray}}
\def\eea{\end{eqnarray}}

\newcommand{\met} {\mbox{$\not\!\!E_T$}}
\begin{document}
\vspace*{4cm}
\title{SUPERSYMMETRY SEARCHES AT THE TEVATRON}

\author{ John Strologas }

\address{Department of Physics and Astronomy\\ University of New Mexico\\
Albuquerque, NM 87131, USA}

\maketitle\abstracts{
We present the current status of searches for supersymmetry performed at the Tevatron accelerator at Fermilab by the CDF and D$\O$ collaborations using luminosity of up to 2.1 fb$^{-1}$.  We focus on searches for charginos, neutralinos, squarks, gluinos and sneutrinos in several supersymmetric scenarios.  No supersymmetric signal is detected and limits on the masses and production cross sections for the supersymmetric particles are set.}

\section{Introduction}

Although the Standard Model (SM) is extremely successful in describing the known elementary particles and their interactions, it is not the final answer to our main questions about the physical world.  Supersymmetry (SUSY), the theory that predicts the existence of a boson for every SM fermion and vice-versa, is one of the promising SM extensions.  With the expense of doubling the known particles and introducing new interaction couplings, SUSY removes the fine tuning of radiative corrections making perturbative quantum field theory valid at arbitrarily high energies, offers the possibility of force unification, incorporates gravity through superstrings and provides a cold dark matter candidate, which is the lightest supersymmetric particle (LSP) if it is stable. 
In this paper, we concentrate on the recent searches at the Tevatron for squarks, gluinos, charginos, neutralinos and sneutrinos.  The analyses assume that the breaking of SUSY is mediated either through gravity (mSUGRA scenario, where the LSP is the neutralino) or through gauge-bosons (GMSB scenario, where the LSP is the gravitino).  Most analyses also assume $R$-parity conservation, according to which the supersymmetric particles must be produced in pairs and the LSP is stable and undetectable, resulting in missing transverse energy ($\met$).

\section{Search for chargino-neutralino production}

\begin{figure}[!]
\begin{minipage}[!]{7cm}  
\begin{center}
\includegraphics[scale=.7]{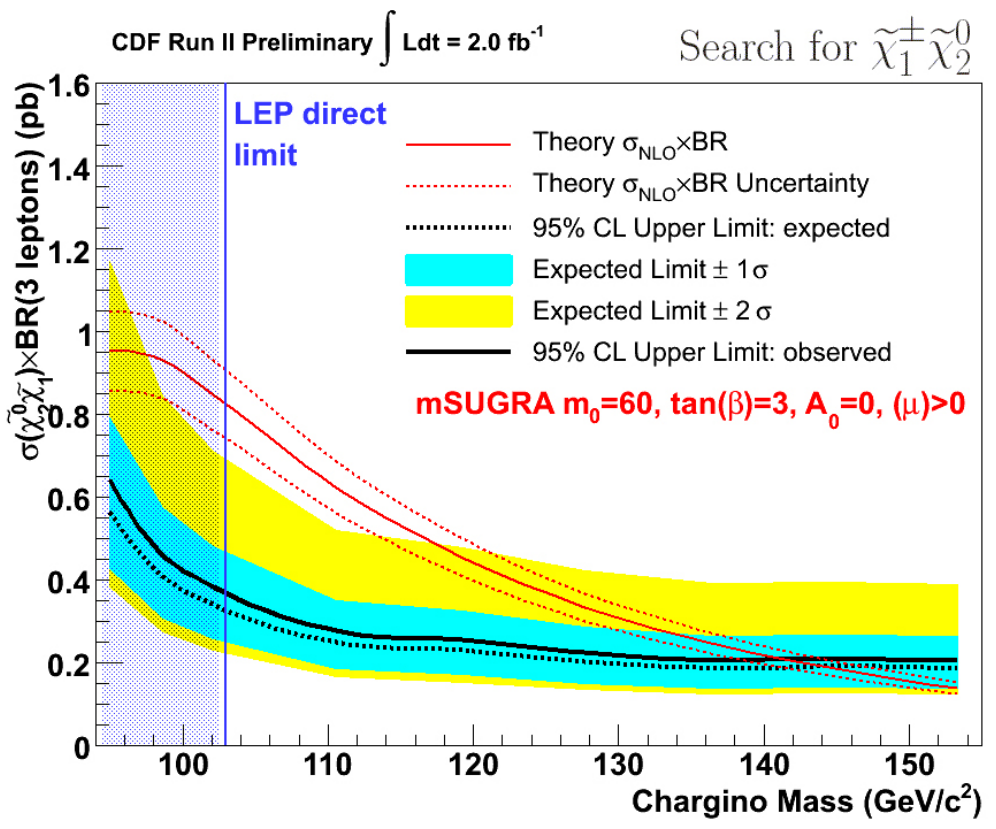}
\caption{Exclusion plot for the CDF chargino-neutralino search. \label{chNCDF}}
\end{center}
\end{minipage}  
\hfill
\begin{minipage}[!]{7cm} 
\begin{center}
\includegraphics[scale=.4]{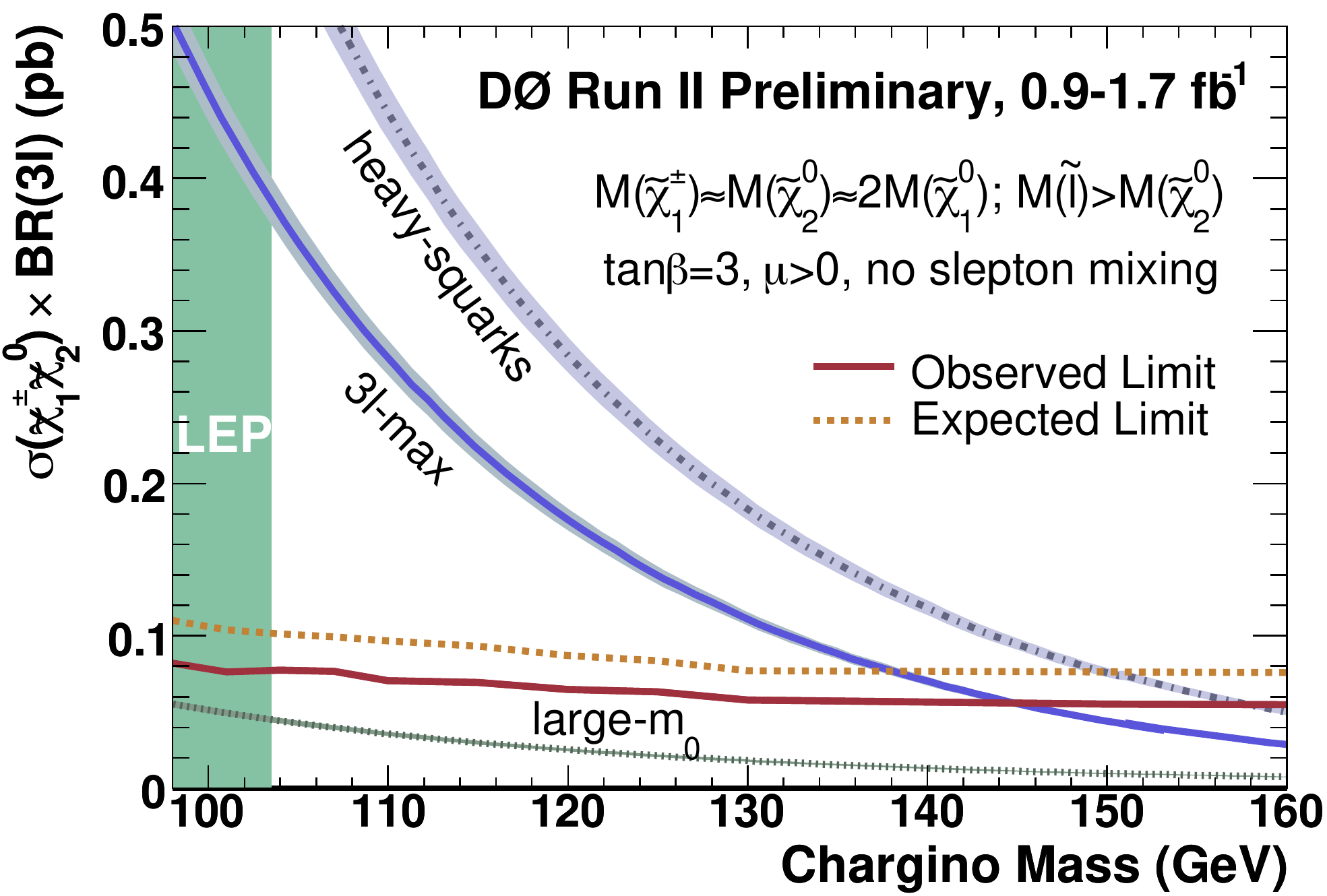}
\caption{Exclusion plot for the D$\O$ chargino-neutralino search. \label{chND0}}
\end{center}
\end{minipage}  
\hfill
\end{figure}

The charginos and the neutralinos are the supersymmetric partners of the gauge and Higgs bosons
and, assuming mSUGRA and $R$-parity conservation, they are produced in pairs and decay to 
the lightest-neutralino which is the LSP.
The golden signature of SUSY at the Tevatron comes from the decay of the lightest chargino ($\tilde{\chi}_1^{\pm}$) and the next-to-lightest neutralino ($\tilde{\chi}^0_2$) to three leptons as well as $\met$ resulting from the final-state lightest neutralino $\tilde{\chi}_1^0$s and neutrino(s).  This signature is characterized by very low SM backgrounds while signal
cross sections of the order of 0.1--1 pb, depending on the mSUGRA parameters, have not been excluded yet. 
CDF has recently completed a 2 fb$^{-1}$ 
trilepton analysis$^1$, which requires three leptons (or two leptons and an isolated track), with
minimum $p_T$ of 15/20, 5/10 and 5/10 GeV/$c$ respectively (depending on the quality of the leptons), $\met$ above 20 GeV, 
low jet multiplicity (for reduction of the top background) and dilepton 
mass above 20 GeV/$c^2$ (with the exclusion of $Z$ mass window), for reduction of 
QCD background, low-mass Drell-Yan (DY) and meson resonances.  
After this selection, the dominant backgrounds are the production of dibosons (including $Z+\gamma$) and the production of DY in association with a ``fake'' third lepton (jet (track) misidentified as electron (muon)).  Overall, $6.4 \pm 1.1$ SM events and $11 \pm 1$ SUSY events ($\tan\beta=3$, $m_0=60$ GeV/$c^2$, $m_{1/2}=190$ GeV/$c^2$, $A_0=0$, $\mu>0$) are expected and 7 events are observed.  The cross section and lightest chargino mass 95\% confidence level (CL) exclusion limits
are 0.2 pb and 140 GeV, for the above mSUGRA point with $m_{1/2}$ varied,
as shown in Figure \ref{chNCDF}.  This is the first pure mSUGRA $\tilde{\chi}_1^{\pm}\tilde{\chi}^0_2$
limit at the Tevatron with no extra assumptions.  
D$\O$ has analyzed 590 pb$^{-1}$ of 
recently collected $ee+\ell$ data$^1$, with 
similar selection requirements, and the results are combined with previous analyses of $\sim 0.9 - 1.1$ fb$^{-1}$.  
Overall, $4 \pm 1$ SM events are expected and 3 are observed.
Figure \ref{chND0} shows the cross section vs. chargino mass exclusion for mSUGRA
with several extra assumptions that include removal of slepton mixing for maximization of acceptance$^1$.

\section{Search for squarks and gluinos}

\begin{figure}[!]
\begin{minipage}[!]{7cm}  
\begin{center}
\includegraphics[scale=.35]{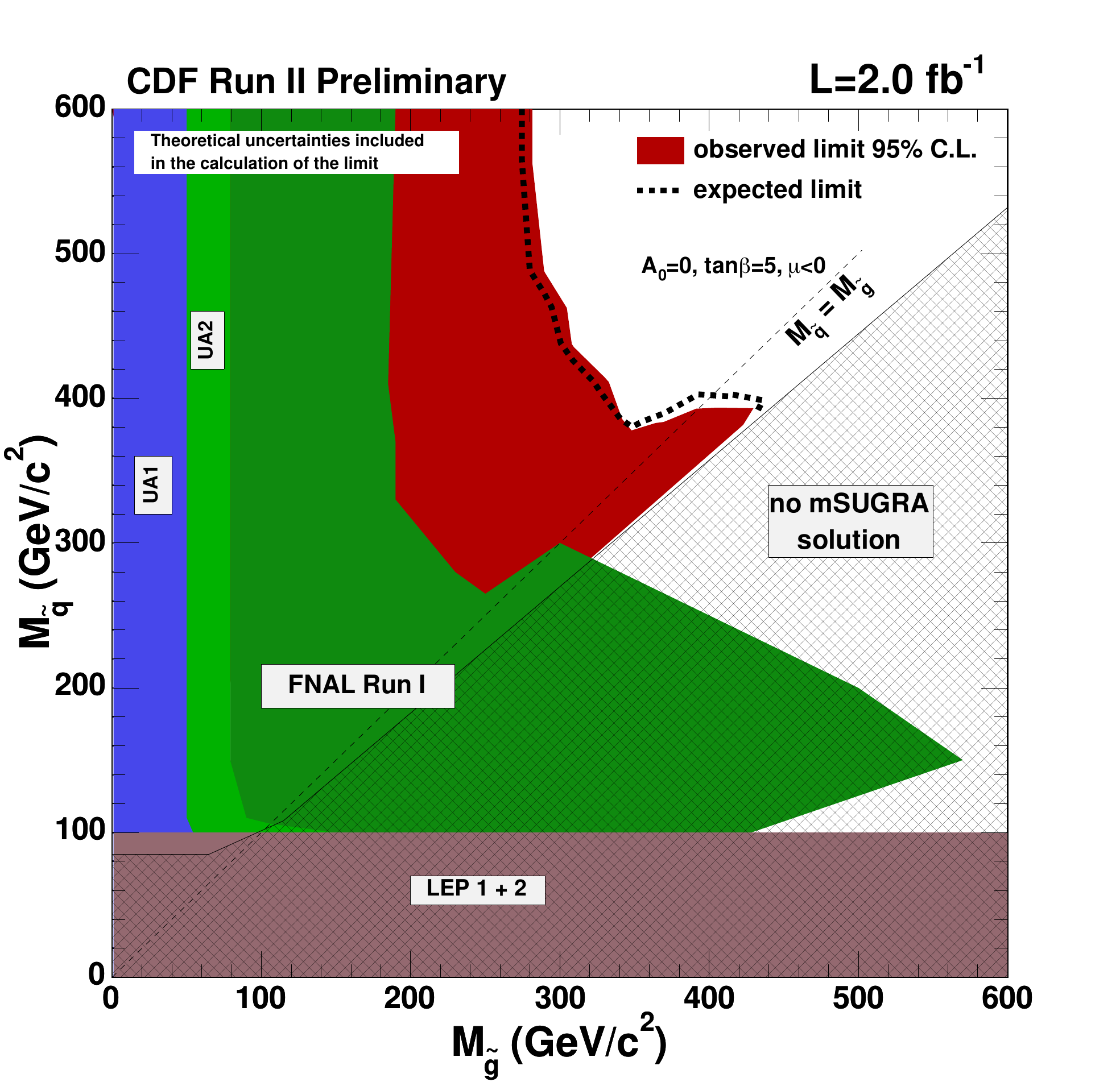}
\caption{The exclusion plot for the CDF squark-gluino analysis. \label{sqGlCDF}}
\end{center}
\end{minipage}  
\hfill
\begin{minipage}[!]{7cm} 
\begin{center}
\includegraphics[scale=.36]{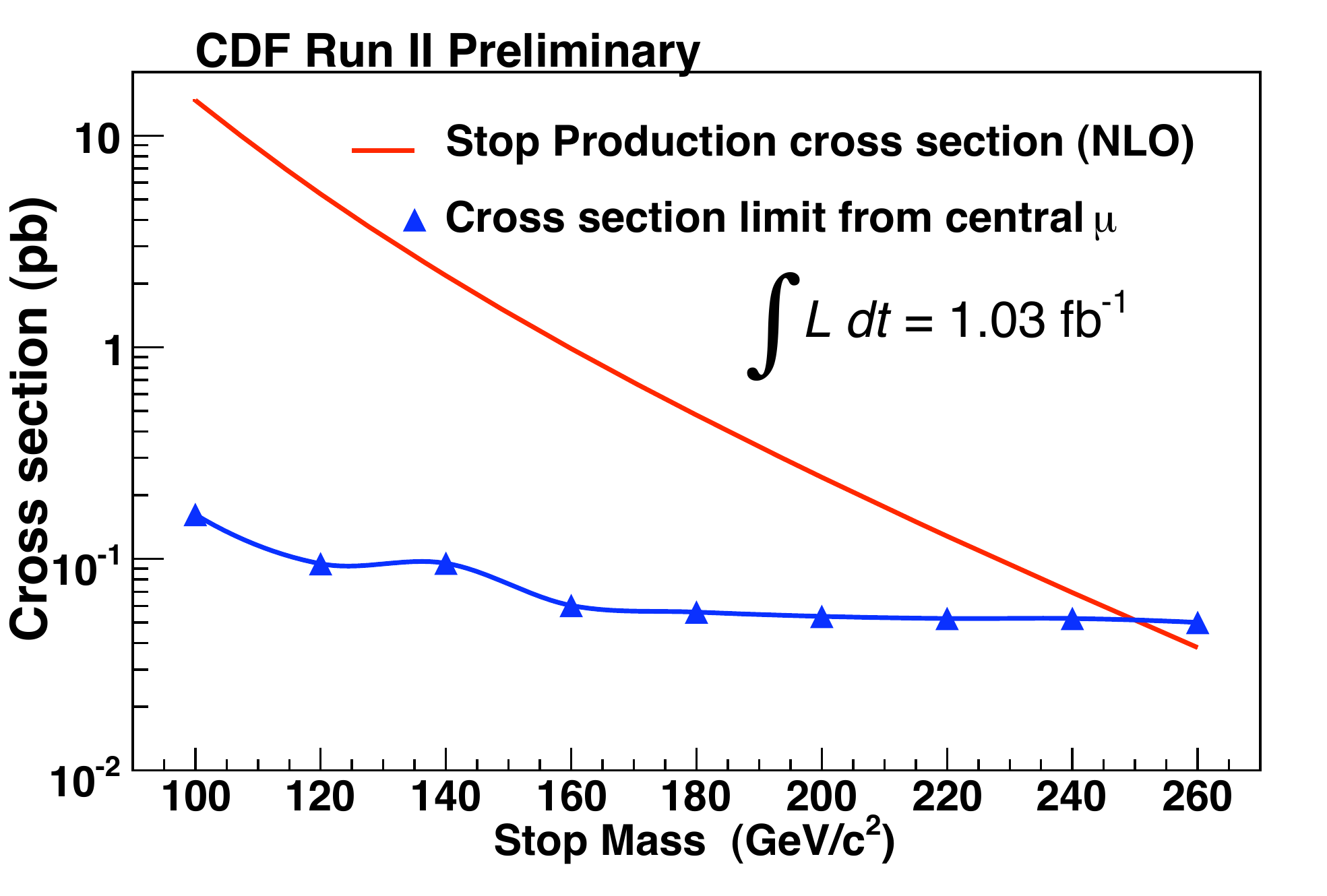}
\caption{The exclusion plot for the CDF stable stop analysis. \label{stabletop}}
\end{center}
\end{minipage}  
\hfill
\end{figure}

The squarks and gluinos are strongly interacting particles 
and therefore could be copiously produced at the Tevatron. 
However, their production cross section is limited by 
their large mass in most of the non-excluded mSUGRA parameter
space.  In addition, they decay into
jets and $\met$, a signature that is dominated by large
QCD multijet background. 
Therefore, careful optimization
of the event selection is critical for the squark-gluinos searches.
Here the light-flavored quarks are considered and
are assumed to be heavy and almost degenerate in mass.
Squarks decay to quarks and the LSP $\tilde{\chi}_1^0$
whereas the gluinos decay to two quarks and the LSP.
The final signature depends on their relative masses
and could be 2 jets and $\met$ (for $M_{\tilde{q}} \ll M_{\tilde{g}}$), or 4 jets+$\met$
($M_{\tilde{q}} \gg M_{\tilde{g}}$) or 3 jets and $\met$ (if $M_{\tilde{q}} \sim M_{\tilde{g}}$).   
The main
backgrounds are $W/Z$+jets, $t\bar{t}$, dibosons 
and, most importantly, the QCD multijet. 
The optimization is done using jet $E_T$, $H_T$ (sum of jet $E_T$) and $\met$,
separately for the three channels, because of the dependence of the 
QCD background and the SUSY signal on jet multiplicity.  The final optimized cuts, 
expected background and observed data events for the three channels and for the two experiments 
can be seen in Table \ref{tab1}.  The CDF $M_{\tilde{q}}$ vs $M_{\tilde{g}}$ 
exclusion plot for $\tan\beta=3$, $A_0=0$, $\mu<0$, and varied $m_0$ and $m_{1/2}$,
can be seen in Figure \ref{sqGlCDF}$^2$.  The D$\O$ exclusion plot is very similar$^3$. 
\begin{table}[t]
\caption{Expected and Observed events in the three jet multiplicity bins
for CDF (2 fb$^{-1}$) and D$\O$ (2.1 fb$^{-1}$).\label{tab1}}
\vspace{0.4cm}
\begin{center} 
\begin{tabular}{|c||c|c|c|c|c|}
\hline
Analysis & $H_T$ cut (GeV) & $\met$ cut (GeV) & Jet $E_T$ cuts (GeV) & SM Background & Data \\ 
\hline
D$\O$ dijet & 325 & 225 & 35,35 & $11^{+3}_{-2}$ & 11\\
D$\O$ trijet & 375 & 175 & 35,35,35 & $11^{+3}_{-2}$ & 9\\
D$\O$ 4-jet & 400 & 100 & 35,35,35,20 & $18^{+6}_{-3}$ & 20\\
\hline
CDF dijet & 330 & 180 & 165,100 & $16 \pm 5$ & 18\\
CDF trijet & 330 & 120 & 140,100,25 & $37 \pm 12$ & 38\\
CDF 4-jet & 280 & 90 & 95,55,55,25 & $48 \pm 17$ & 45\\
\hline
\end{tabular}
\end{center}
\end{table}

One of the stop quarks may be much lighter than the other squarks 
and it may not be able to decay to the lightest neutralino and a quark.
If its mass is such that $m_c + m_{\tilde{\chi}_1^0}<m_{\tilde{t}}<m_b+m_W+m_{\tilde{\chi}_1^0}$ 
then it will decay to a charm quark through a flavor-changing loop process.  
A 1~fb$^{-1}$ analysis of stop to charm is completed by D$\O$$^4$.  Exactly
two jets (with $E_T$ greater than 20 and 40 GeV) are required with proper angular separation from the 
$\met$ for reduction of the QCD and $W$+jet background.  Heavy-flavor tagging 
is implemented with the use of a neural network.  Using the $H_T$/$\met$-optimized cuts 
at 140/70 GeV, $64\pm 3$ SM events are expected and 66 are observed.  
This result corresponds to exclusion of stop masses below 149 GeV/$c^2$ 
for neutralino masses below 63 GeV/$c^2$ at 95\% CL.

If the stop is too light to decay to charm, it will not
decay inside the detector and it will look like a CHarged Massive Particle (CHAMP).
The CDF collaboration performed a search$^5$ for stable CHAMP-like
top, which could be reconstructed as a heavy muon with 
possible energy deposition in the calorimeter.  The measurement is performed
by reconstructing the mass of the slow massive particle using its $\beta$,
measured with the Time of Flight system and the momentum of its 
corresponding track.  The backgrounds come from cosmic rays and 
multiple interactions both of which would create fake delayed signals.
In the 1 fb$^{-1}$ analysis, $4.7 \pm 0.3$ SM events are expected and 4 are observed.  
A stable stop is excluded below 250 GeV/$c^2$ at 95\% CL.
  
\section{$\bm R$-parity violation signals}

\begin{figure}[!]
\begin{minipage}[!]{7cm}  
\begin{center}
\includegraphics[scale=.4]{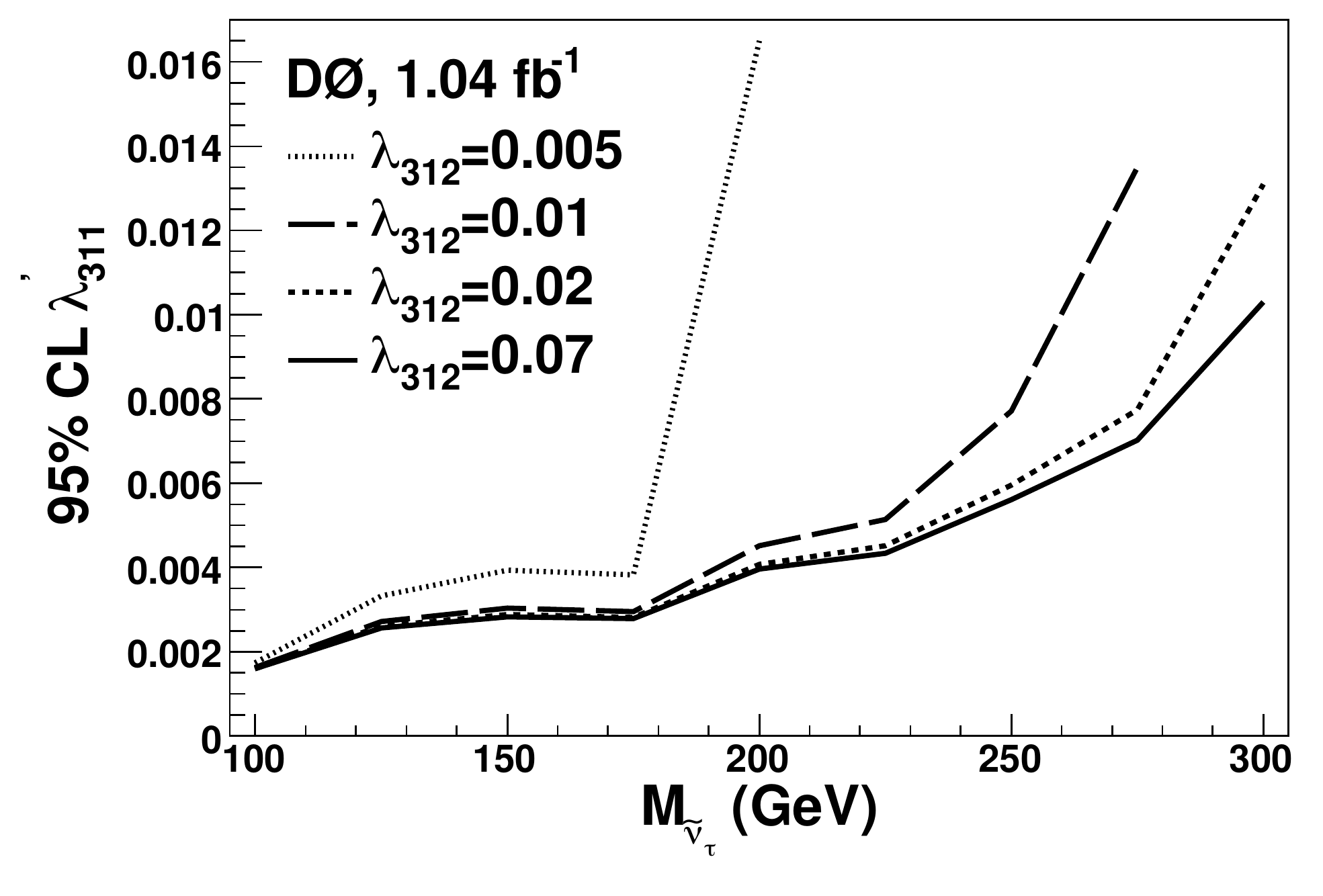}
\caption{The exclusion plot for the D$\O$ sneutrino analysis. \label{sneutrino}}
\end{center}
\end{minipage}  
\hfill
\begin{minipage}[!]{7cm} 
\begin{center}
\includegraphics[scale=.4]{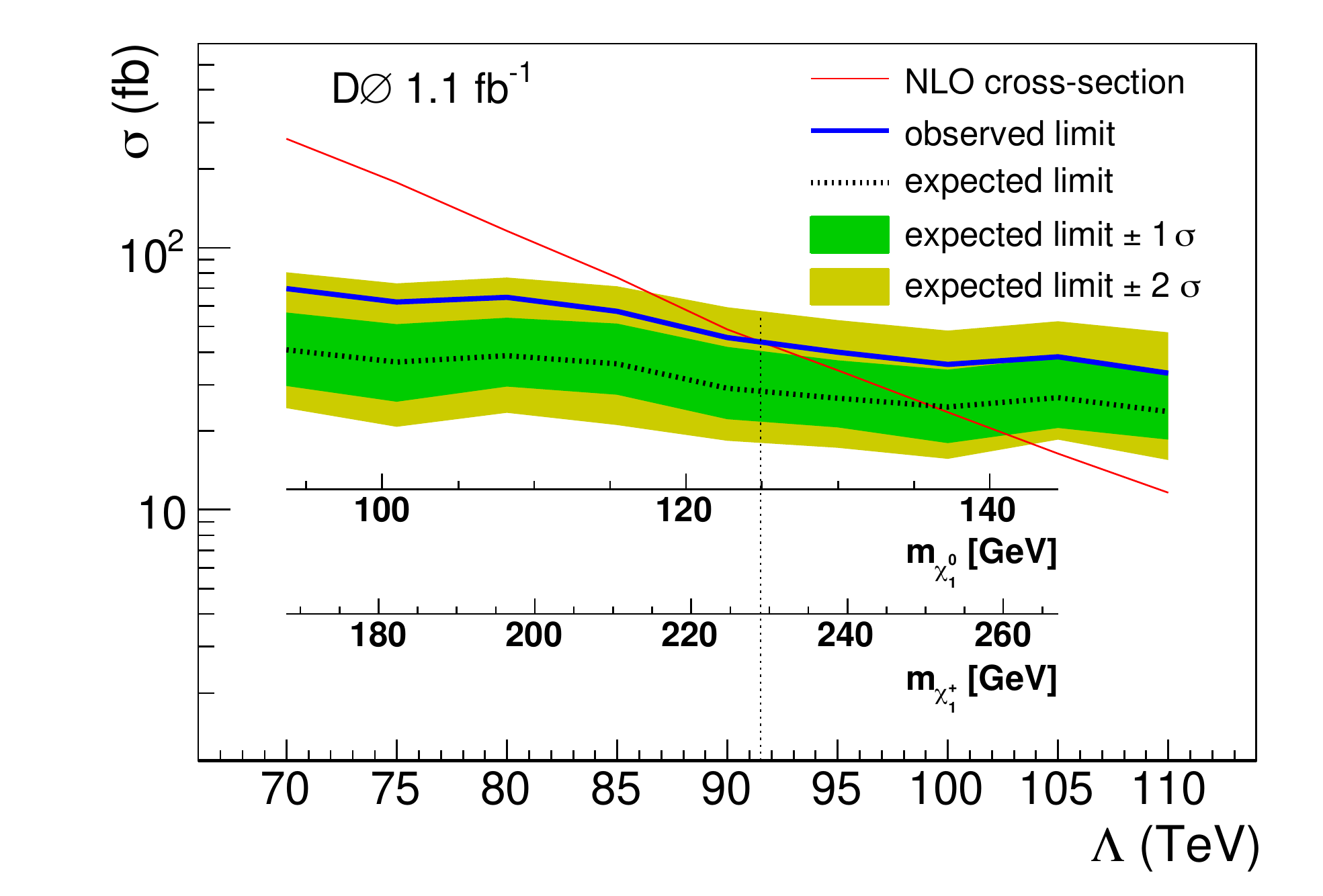}
\caption{The exclusion plot for the D$\O$ GMSB diphoton analysis. \label{diphoton}}
\end{center}
\end{minipage}  
\hfill
\end{figure}

If we do not demand $R$-parity conservation then 
lepton- and baryon-number violating terms are allowed 
in the soft SUSY-breaking sector of the lagrangian. 
D$\O$ performed a 1~fb$^{-1}$ analysis$^6$ looking for lepton-number-violating
sneutrinos decaying to $e+\mu$. An electron ($30<E_T<100$~GeV), a muon ($p_T>25$ GeV/$c$)
and $\met<15$ GeV and no extra leptons are required.
The main backgrounds come from $Z/\gamma\rightarrow \tau\tau$ and 
$WW$ production.  Overall, $59\pm 5$ SM events are expected and 68
observed.  Figure \ref{sneutrino} shows the 95\% CL exclusion limits 
for $\lambda'$ as a function of the sneutrino mass for several $\lambda$ values,
where $\lambda$ and $\lambda'$ are the lepton-violating couplings.
CDF has performed a search for $R$-parity violation with multileptons
at 346 fb$^{-1}$, using at least 3 leptons (electrons or muons)$^7$.  
The lower 95\% CL mass limits for $\tilde{\chi}_1^0$ 
and $\tilde{\chi}_1^{\pm}$ are set to $98 - 110$ and $185 - 202$ GeV/$c^2$ respectively,
depending on the $\lambda$ assumptions.
\section{Search for GMSB neutralinos}
In the GMSB scenario, the pair production of 
neutralinos would result to 2 gravitinos (LSPs) and 2 photons.
A 1.1~fb$^{-1}$ D$\O$ analysis$^8$ searched for two prompt photons above 25 GeV and $\met>$30 or 60 GeV,
with angular separation between any present jet and the $\met$.
The background comes from processes with no real diphotons
($W\gamma$, $W+{\rm jet}$, $t\bar{t}$, QCD-multijet) and at 
a lesser degree from processes with real diphotons
($W\gamma\gamma$ and $Z\gamma\gamma$). 
Overall, $11\pm 1$ SM events for $\met>30$ and $1.6\pm 0.4$ SM events
for $\met>60$ are expected and 16 and 3 are observed respectively.
Figure \ref{diphoton} shows the cross section limit as a function of the GMSB $\Lambda$
parameter and the $\tilde{\chi}_1^{+}$ and $\tilde{\chi}_1^{0}$ masses.
\section*{References}
\begin{tabbing}
1. CDF public note \#9176, D$\O$ public note \#5464. \hspace{0.8cm}\= 2. CDF public note \#9229.\\
3. D$\O$, Phys.\ Lett.\ B {\bf 660} 449 (2008). \> 4. D$\O$, arXiv:0803.2263v1 [hep-ex].\\
5. CDF public note \#8701. \> 6. D$\O$, arXiv:0711.3207v2 [hep-ex].\\
7. CDF, Phys.\ Rev.\ Lett.\ {\bf 98}, 131804 (2007). \> 8. D$\O$, Phys.\ Lett.\ B {\bf 659} 856 (2008).
\end{tabbing}

\end{document}